\documentclass[twocolumn,floatfix,showpacs,preprintnumbers,amsmath,amssymb,superscriptaddress,prl]{revtex4-1}
\usepackage[latin9]{inputenc}
\usepackage{geometry}
\usepackage[active]{srcltx}
\usepackage{xcolor}
\usepackage{array}
\usepackage{rotating}
\usepackage{float}
\usepackage{rotfloat}
\usepackage{textcomp}
\usepackage{multirow}
\usepackage{amsmath}
\usepackage{amssymb}
\usepackage{graphicx}
\usepackage{setspace}

\makeatletter

\DeclareFontEncoding{LGR}{}{}
\DeclareRobustCommand{\greektext}{%
  \fontencoding{LGR}\selectfont\def\encodingdefault{LGR}}
\DeclareRobustCommand{\textgreek}[1]{\leavevmode{\greektext #1}}
\ProvideTextCommand{\~}{LGR}[1]{\char126#1}

\providecommand{\tabularnewline}{\\}

\usepackage{graphics}\usepackage{subfigure}\usepackage{longtable}\usepackage{pstricks}\usepackage{dcolumn}\usepackage{bm}

\makeatother

\begin{document}

\title{Kagome quantum spin systems in the atacamite family}

\author{Pascal Puphal}

\affiliation{Physikalisches Institut, Goethe-Universität Frankfurt, 60438 Frankfurt
am Main, Germany}

\author{Katharina M. Zoch}

\affiliation{Physikalisches Institut, Goethe-Universität Frankfurt, 60438 Frankfurt
am Main, Germany}

\author{Joy Désor}

\affiliation{Physikalisches Institut, Goethe-Universität Frankfurt, 60438 Frankfurt
am Main, Germany}

\author{Michael Bolte}

\affiliation{Institut f\"ur Organische Chemie der Universit\"at Frankfurt, 60439
Frankfurt am Main, Germany}

\author{Cornelius Krellner}

\affiliation{Physikalisches Institut, Goethe-Universität Frankfurt, 60438 Frankfurt
am Main, Germany}
\begin{abstract}
We present the hydrothermal synthesis, as well as structural and chemical
analysis of single crystals on the compounds EuCu$_{3}$(OH)$_{6}$Cl$_{3}$,
Zn$_{x}$Cu$_{4-x}$(OH)$_{6}$(NO$_{3}$)$_{2}$ and haydeeite, MgCu$_{3}$(OH)$_{6}$Cl$_{2}$
all arising from the atacamite family. Magnetic and specific-heat
measurements down to 1.8~K are carried out for these systems. EuCu$_{3}$(OH)$_{6}$Cl$_{3}$
has a frustrated antiferromagnetic Cu$^{2+}$ ground state with order
at 15\,K, a strong anisotropy and increased magnetization from Van
Vleck paramagnetic Eu$^{3+}$ contributions. ZnCu$_{3}$(OH)$_{6}$(NO$_{3}$)$_{2}$
reveals antiferromagnetic order at 9\,K and measurements on haydeeite
single crystals confirm the ferromagnetic order at 4.2\,K with the
easy axis within the kagome plane. These results prove that the atacamite
family presents a broad class of materials with interesting magnetic
ground states.
\end{abstract}
\maketitle

\section{Introduction}

The discovery of the first reported prototype with magnetic Cu$^{2+}-$ions
arranged on a kagome layer presenting a quantum spin liquid (QSL \citep{Balens(2010)})
in herbertsmithite (ZnCu$_{3}$(OH)$_{6}$Cl$_{2}$) \citep{Shores(2005)},
triggered enormous interest for this novel ground state. \textcolor{black}{The
dominant Cu-O-Cu antiferromagnetic superexchange of J $=$ -197\,K
\citep{Helton(2007)} in the kagome plane is strongly frustrated due
to the geometrical arrangement of the ions. Considering the possibility
of variable Zn for Cu substitution in Zn$_{x}$Cu$_{4-x}$(OH)$_{6}$Cl$_{2}$,
one can influence the magnetic properties of the compounds: increasing
the Zn in between these kagome layers leads to their magnetic decoupling
and thus to a suppression of the magnetic order \citep{Mendels(2007)}.
For the composition ZnCu$_{3}$(OH)$_{6}$Cl$_{2}$ ($x=1$), no magnetic
long-range order is registered down to T = 50~mK \citep{Helton(2007)}.}
Later on, it has been realized that some Zn-Cu antisite disorder (up
to 15\% in terms of occupation parameters) is always present in herbertsmithite
\citep{Vries(2012),Han(2016)}, which has led to the search for further
kagome materials, with less amount of structural disorder. The atacamite
family of compounds presents a rich field of different substitution
possibilities \citep{Norman(2016)} based on the three basic polymorphs
of Cu$_{2}$(OH)$_{3}$Cl: atacamite, clinoatacamite and botallackite,
allowing substitutions both on the Cu and Cl place. Cationic substitutions
would dilute the concentration of Cu$^{2+}$ ions between and in the
kagome planes, while those on the anionic sites would have a direct
impact on the spatial separation of these active planes. Furthermore,
it was proposed that substituting trivalent ions on the divalent Zn/Cu
ions will lead to a correlated Dirac-kagome metal combining Dirac
electrons, strong interactions, and frustrated magnetism \citep{Mazin(2014)}.
However, experimentally it was found that the structure presents charge
balancing with additional Cl$^{-}$ rather than electron doping, as
seen for the nonmagnetic Y \citep{Puphal2017,Sun(2016)} and the magnetic
Nd, Gd and Sm \citep{Sun(2017)} ions. 

In table \ref{tab:List} we present the whole class of M$_{x}$Cu$_{4-x}$(OH)$_{6}$X$_{2}$
with detailed information on the structure type and magnetism. The
atom type of substitution is highlighted by a color and its crystal
ionic radius is given in the fourth column. The given space group
in the fifth column enables a division into two main structural variants:
the herbertsmithite R$\bar{3}$m type structure containing ABC stacked
kagome layers with intermediate layers of the substituent and the
kapellasite type with AA stacked layers, where the substituent is
in the kagome layer in the center of the star (see figure \ref{Eu structure}).
Other groups are the basic structures with possible substitutions
on the halide site, then the crossover paratacamite over herbertsmithite
to the substitutions of $x=2$ leading with nonmagnetic ions to one
dimensional chain systems and finally the $x=4$ entirely substituted
variants. The last type shows the structural variance of this family
and the possibility of undescribed intermediate candidates. We want
to highlight the work of reference \citep{M2(OH)3Cl} which already
shows a rich amount of variants only for the M$_{2}$(OH)$_{3}$(Cl,Br,I)
candidates. Besides the shown atacamite type a ``kapellasite-type''
structure called $\alpha$ variant was found by Oswald and Feitknecht
\citep{M2(OH)3Cl} for Co, Ni, Mg, Fe, Mn with $x=4$ substitution.
As well as paratacamite variants of Co/ Fe and botallackite types
for Br/ I both for Cu, Co, Ni, Fe as well as Mn.

The columns (6-9) give details about the magnetic properties and is
followed by some structural details of the Cu ions (10-12), where
Cu$_{K}$ stands for the shortest copper distances in the kagome plane
and Cu$_{inter}$ for the distances between these layers ignoring
Cu atoms in between. The column number 12 contains the bonding angle
of Cu-O-Cu where the largest and thus dominating J is chosen. In the
last two columns references are given. So far the only kagome systems
of this family presenting no magnetic order down to mK temperatures
are both ZnCu$_{3}$(OH)$_{6}$Cl$_{2}$ herbertsmithite and kapellasite,
ZnCu$_{3}$(OH)$_{6}$FBr called Zn-barlowite and MgCu$_{3}$(OH)$_{6}$Cl$_{2}$
tondiite which all have a strong Zn-Cu or Mg-Cu exchange. Other quantum
spin liquid candidates of the family are ones presenting isolated
trimers found in \textcolor{black}{SrCu(OH)$_{3}$Cl }\citep{SrCu(OH)3Cl}\textcolor{black}{{}
and ZnCu$_{3}$(OH)$_{6}$SO$_{4}$ }\citep{namuwite}, which so far
only exist as polycristals.

In addition to these compounds we present another example of the P$\bar{3}$m1
structure with a trivalent Eu$^{3+}$-ion realized in EuCu$_{3}$(OH)$_{6}$Cl$_{3}$.
As well as an example of an antiferromagnetic system with no strong
frustration effects, observed in ZnCu$_{3}$(OH)$_{6}$(NO$_{3}$)$_{2}$
and the successful single crystal growth of the ferromagnetic haydeeite
MgCu$_{3}$(OH)$_{6}$Cl$_{2}$. 

\onecolumngrid

\begin{sidewaystable}
\begin{spacing}{1.1}
\begin{tabular}{lllll|llll|lll|l|l}
{\footnotesize{}Type} & {\footnotesize{}Mineral } & {\footnotesize{}Formula} & {\footnotesize{}r$_{\text{crys}}${[}pm{]}} & {\footnotesize{}Point } & {\footnotesize{}Magnetic} & {\footnotesize{}Magnetic } & {\footnotesize{}$\Theta_{\text{W}}$ {[}K{]}} & {\footnotesize{}f} & {\footnotesize{}Cu$\text{\ensuremath{_{K}}}$ {[}$\textrm{Å}${]}} & {\footnotesize{}Cu$_{\text{inter}}$ {[}$\textrm{Å}${]}} & {\footnotesize{}$\angle$(Cu1-O1} & {\footnotesize{}Reference} & {\footnotesize{}sc growth}\tabularnewline
 & {\footnotesize{}name} &  & {\footnotesize{} \citep{radii2,radii,anion radii}} & {\footnotesize{}group} & {\footnotesize{} lattice} & {\footnotesize{}order {[}K{]}} &  &  &  &  & {\footnotesize{}-Cu1) {[}\textdegree {]}} & {\footnotesize{}(mag, struc)} & \tabularnewline
\hline 
\multirow{3}{*}{\begin{turn}{90}
{\footnotesize{}parent}
\end{turn}} & {\footnotesize{}atacamite} & {\footnotesize{}Cu$_{2}$(OH)$_{3}$Cl} & {\footnotesize{}72} & {\footnotesize{}Pnma} & {\footnotesize{}P} & {\footnotesize{}9} & {\footnotesize{}-125} & {\footnotesize{}14} & {\footnotesize{}3.432} & {\footnotesize{}5.464} & {\footnotesize{}124.39} & {\footnotesize{}\citep{atacamite}} & {\footnotesize{}-}\tabularnewline
 & {\footnotesize{}clinoatacamite} & {\footnotesize{}Cu$_{2}$(OH)$_{3}$Cl} & {\footnotesize{}72} & {\footnotesize{}P2$_{1}$/n} & {\footnotesize{}P} & {\footnotesize{}6.5} & {\footnotesize{}-200} & {\footnotesize{}30} & {\footnotesize{}3.410} & {\footnotesize{}5.023} & {\footnotesize{}120.21} & {\footnotesize{}\citep{clinoatacamite}} & {\footnotesize{}\citep{clino}}\tabularnewline
 & {\footnotesize{}botallackite} & {\footnotesize{}(Ni/Co/}\textcolor{violet}{\footnotesize{}Fe}{\footnotesize{}/}\textcolor{gray}{\footnotesize{}Mn}{\footnotesize{}/Cu)$_{2}$-} & \textit{\footnotesize{}69/ 74.5/ }\textit{\textcolor{violet}{\footnotesize{}75}}\textit{\footnotesize{}/
}\textit{\textcolor{gray}{\footnotesize{}83}} & {\footnotesize{}P2$_{1}$/m} & {\footnotesize{}T} & {\footnotesize{}7.2/}\textcolor{teal}{\footnotesize{}10}{\footnotesize{}/}\textcolor{orange}{\footnotesize{}14} & {\footnotesize{}0.41} & {\footnotesize{}0} & {\footnotesize{}3.059} & {\footnotesize{}5.716} & {\footnotesize{}107.91} & {\footnotesize{}\citep{atacamite,M2(OH)3Cl}} & {\footnotesize{}-}\tabularnewline
 &  & {\footnotesize{}(OH)$_{3}$(Cl/}\textcolor{teal}{\footnotesize{}Br}{\footnotesize{}/}\textcolor{orange}{\footnotesize{}I}{\footnotesize{})} & {\footnotesize{}172/ }\textcolor{teal}{\footnotesize{}188}{\footnotesize{}/
}\textcolor{orange}{\footnotesize{}210} &  &  &  &  &  &  &  &  &  & \tabularnewline
 & {\footnotesize{}brochantite} & {\footnotesize{}Cu$_{4}$(OH)$_{6}$}\textcolor{purple}{\footnotesize{}SO$_{4}$} & \textcolor{purple}{\footnotesize{}258} & {\footnotesize{}P2$_{1}$/n} & {\footnotesize{}'1D'} & {\footnotesize{}6.3} & {\footnotesize{}-90} & {\footnotesize{}14} & {\footnotesize{}3.005} & {\footnotesize{}5.062} & {\footnotesize{}124.5} & {\footnotesize{}\citep{brochantite}} & {\footnotesize{}-}\tabularnewline
 & {\footnotesize{}rouaite} & {\footnotesize{}Cu$_{2}$(OH)$_{3}$}\textcolor{violet}{\footnotesize{}$\left(\text{NO}_{3}\right)$} & \textcolor{violet}{\footnotesize{}179} & {\footnotesize{}P2$_{1}$} & {\footnotesize{}T} & {\footnotesize{}11} & {\footnotesize{}-12} & {\footnotesize{}1} & {\footnotesize{}3.05} & {\footnotesize{}6.929} & {\footnotesize{}110.33} & {\footnotesize{}\citep{claringbullite,ROUAITE}} & {\footnotesize{}-}\tabularnewline
 & {\footnotesize{}claringbullite} & {\footnotesize{}Cu$_{4}$(OH)$_{6}$Cl}\textcolor{pink}{\footnotesize{}F} & \textcolor{pink}{\footnotesize{}126} & {\footnotesize{}P6$_{3}$/mmc} & {\footnotesize{}P} & {\footnotesize{}17} & {\footnotesize{}-33} & {\footnotesize{}2} & {\footnotesize{}3.337} & {\footnotesize{}2.737} & {\footnotesize{}117.56} & {\footnotesize{}\citep{claringbullite,Claringbull}} & {\footnotesize{}-}\tabularnewline
 & {\footnotesize{}barlowite} & {\footnotesize{}Cu$_{4}$(OH)$_{6}$}\textcolor{teal}{\footnotesize{}Br}\textcolor{pink}{\footnotesize{}F} & \textcolor{teal}{\footnotesize{}188}{\footnotesize{}, }\textcolor{pink}{\footnotesize{}126} & {\footnotesize{}P6$_{3}$/mmc} & {\footnotesize{}P} & {\footnotesize{}15} & {\footnotesize{}-136} & {\footnotesize{}9} & {\footnotesize{}3.339} & {\footnotesize{}2.757} & {\footnotesize{}117.21} & {\footnotesize{}\citep{barlowite}} & {\footnotesize{}\citep{barlow crys}}\tabularnewline
\hline 
 & {\footnotesize{}paratacamite} & \textcolor{brown}{\footnotesize{}Zn}{\footnotesize{}$_{x}$Cu$_{4-x}$(OH)$_{6}$Cl$_{2}$} & \textcolor{brown}{\footnotesize{}73} & {\footnotesize{}R$\bar{3}$m} & {\footnotesize{}K} & {\footnotesize{}6 } & {\footnotesize{}-231} & {\footnotesize{}38.5} & {\footnotesize{}3.416} & {\footnotesize{}3.110} & {\footnotesize{}122.28} & {\footnotesize{}\citep{Shores(2005),para} $x=0.5/0.25$} & {\footnotesize{}\citep{clino}}\tabularnewline
\hline 
\multirow{7}{*}{\begin{turn}{90}
{\footnotesize{}``herbertsmithite-type''}
\end{turn}} & {\footnotesize{}-} & \textcolor{green}{\footnotesize{}Ga}{\footnotesize{}$_{0.8}$Cu$_{3.2}$(OH)$_{6}$Cl$_{2}$} & \textcolor{green}{\footnotesize{}62} & {\footnotesize{}R$\bar{3}$m} & {\footnotesize{}K} & {\footnotesize{}5} & {\footnotesize{}-256} & {\footnotesize{}55} & {\footnotesize{}3.421} & {\footnotesize{}5.066} & {\footnotesize{}118.88} & {\footnotesize{}\citep{Ga}} & {\footnotesize{}-}\tabularnewline
 & \textit{\footnotesize{}gillardite} & \textit{\footnotesize{}NiCu$_{3}$(OH)$_{6}$Cl$_{2}$} & \textit{\footnotesize{}69} & \textit{\footnotesize{}R$\bar{3}$m} & \textit{\footnotesize{}K} & \textit{\footnotesize{}6} & \textit{\footnotesize{}-100} & \textit{\footnotesize{}17} & \textit{\footnotesize{}3.418} & \textit{\footnotesize{}5.020} & \textit{\footnotesize{}119.19} & \textit{\footnotesize{}\citep{nico,gillardite}} & {\footnotesize{}-}\tabularnewline
 & {\footnotesize{}tondiite} & \textcolor{blue}{\footnotesize{}Mg}{\footnotesize{}Cu$_{3}$(OH)$_{6}$Cl$_{2}$} & \textcolor{blue}{\footnotesize{}72} & {\footnotesize{}R$\bar{3}$m} & {\footnotesize{}K} & {\footnotesize{}-} & {\footnotesize{}-300} & {\footnotesize{}$\infty$} & {\footnotesize{}3.416} & {\footnotesize{}5.054} & {\footnotesize{}119.05} & {\footnotesize{}\citep{tondiite}} & {\footnotesize{}\citep{tondiite}}\tabularnewline
 & {\footnotesize{}herbertsmithite} & \textcolor{brown}{\footnotesize{}Zn}{\footnotesize{}Cu$_{3}$(OH)$_{6}$Cl$_{2}$} & \textcolor{brown}{\footnotesize{}73} & {\footnotesize{}R$\bar{3}$m} & {\footnotesize{}K} & {\footnotesize{}-} & {\footnotesize{}-300} & {\footnotesize{}$\infty$} & {\footnotesize{}3.416} & {\footnotesize{}5.087} & {\footnotesize{}118.92} & {\footnotesize{}\citep{Shores(2005),herbert}} & {\footnotesize{}\citep{clino}}\tabularnewline
 & {\footnotesize{}-} & \textcolor{brown}{\footnotesize{}Zn}{\footnotesize{}Cu$_{3}$(OH)$_{6}$}\textcolor{teal}{\footnotesize{}Br}\textcolor{pink}{\footnotesize{}F} & \textcolor{brown}{\footnotesize{}73, }\textcolor{teal}{\footnotesize{}188}{\footnotesize{},
}\textcolor{pink}{\footnotesize{}126} & {\footnotesize{}P6$_{3}$/mmc} & {\footnotesize{}K} & {\footnotesize{}-} & {\footnotesize{}-205} & {\footnotesize{}$\infty$} & {\footnotesize{}3.337} & {\footnotesize{}4.660} & {\footnotesize{}117.01} & {\footnotesize{}\citep{Znbarlowite}} & {\footnotesize{}-}\tabularnewline
 & \textit{\footnotesize{}leverettite} & \textit{\footnotesize{}CoCu$_{3}$(OH)$_{6}$Cl$_{2}$} & \textit{\footnotesize{}74.5} & \textit{\footnotesize{}R$\bar{3}$m} & \textit{\footnotesize{}K} & \textit{\footnotesize{}3} & \textit{\footnotesize{}-40} & \textit{\footnotesize{}13} & \textit{\footnotesize{}3.421} & \textit{\footnotesize{}5.096} & \textit{\footnotesize{}119.19} & \textit{\footnotesize{}\citep{nico,leverettite}} & {\footnotesize{}-}\tabularnewline
 & {\footnotesize{}-} & \textit{\textcolor{cyan}{\footnotesize{}Cd}}\textit{\footnotesize{}Cu$_{3}$(OH)$_{6}$Cl$_{2}$} & \textit{\textcolor{cyan}{\footnotesize{}95}} & \textit{\footnotesize{}P2$_{1}$/n} & \textit{\footnotesize{}K} & \textit{\footnotesize{}-} & \textit{\footnotesize{}-150} & \textit{\footnotesize{}75} & \textit{\footnotesize{}3.015} & \textit{\footnotesize{}5.758} & \textit{\footnotesize{}108.50} & \textit{\footnotesize{}\citep{CdCl2}} & {\footnotesize{}-}\tabularnewline
\hline 
\multirow{11}{*}{\begin{turn}{90}
{\footnotesize{}``kapellasite-type''}
\end{turn}} & {\footnotesize{}haydeeite} & \textcolor{blue}{\footnotesize{}Mg}{\footnotesize{}Cu$_{3}$(OH)$_{6}$Cl$_{2}$} & \textcolor{blue}{\footnotesize{}72} & {\footnotesize{}P$\bar{3}$m1} & {\footnotesize{}K} & {\footnotesize{}4.3 (FM)} & {\footnotesize{}28} & {\footnotesize{}7} & {\footnotesize{}3.137} & {\footnotesize{}5.750} & {\footnotesize{}104.98} & {\footnotesize{}\citep{haydeeite,haydeeite struc}} & {\footnotesize{}this paper}\tabularnewline
 & {\footnotesize{}kapellasite} & \textcolor{brown}{\footnotesize{}Zn}{\footnotesize{}Cu$_{3}$(OH)$_{6}$Cl$_{2}$} & \textcolor{brown}{\footnotesize{}73} & {\footnotesize{}P$\bar{3}$m1} & {\footnotesize{}K} & {\footnotesize{}-} & {\footnotesize{}9,5} & {\footnotesize{}$\infty$} & {\footnotesize{}3.150} & {\footnotesize{}5.733} & {\footnotesize{}105.84} & {\footnotesize{}\citep{haydeeite,kapellasite}} & {\footnotesize{}-}\tabularnewline
 & {\footnotesize{}-} & \textcolor{brown}{\footnotesize{}Zn}{\footnotesize{}Cu$_{3}$(OH)$_{6}$}\textcolor{violet}{\footnotesize{}$\left(\text{NO}_{3}\right)_{2}$} & \textcolor{brown}{\footnotesize{}73, }\textcolor{violet}{\footnotesize{}179} & {\footnotesize{}P2$_{1}$} & {\footnotesize{}K?} & {\footnotesize{}7.5} & {\footnotesize{}0} & {\footnotesize{}0} & {\footnotesize{}3.065} & {\footnotesize{}6.927} & {\footnotesize{}104.07} & {\footnotesize{}this paper} & {\footnotesize{}this paper}\tabularnewline
 & \textit{\footnotesize{}misakiite} & \textit{\textcolor{gray}{\footnotesize{}Mn}}\textit{\footnotesize{}Cu$_{3}$(OH)$_{6}$Cl$_{2}$} & \textit{\textcolor{gray}{\footnotesize{}83}} & \textit{\footnotesize{}P$\bar{3}$m1} & \textit{\footnotesize{}T} & \textit{\footnotesize{}10} & \textit{\footnotesize{}-25} & \textit{\footnotesize{}3} & \textit{\footnotesize{}3.208} & \textit{\footnotesize{}5.710} & \textit{\footnotesize{}108.00} & \textit{\footnotesize{}\citep{misakiite}} & {\footnotesize{}-}\tabularnewline
 & {\footnotesize{}-} & \textcolor{purple}{\footnotesize{}Y}{\footnotesize{}Cu$_{3}$(OH)$_{6}$Cl$_{3}$} & \textcolor{purple}{\footnotesize{}90} & {\footnotesize{}P$\bar{3}$m1} & {\footnotesize{}K} & {\footnotesize{}12} & {\footnotesize{}-100} & {\footnotesize{}8.33} & {\footnotesize{}3.250} & {\footnotesize{}5.618} & {\footnotesize{}117.36} & {\footnotesize{}\citep{Sun(2016)}} & {\footnotesize{}-}\tabularnewline
 & {\footnotesize{}-} & \textcolor{purple}{\footnotesize{}Y}{\footnotesize{}$_{3}$Cu$_{9}$(OH)$_{19}$Cl$_{8}$} & \textcolor{purple}{\footnotesize{}90} & {\footnotesize{}R$\bar{3}$} & {\footnotesize{}K} & {\footnotesize{}2.2} & {\footnotesize{}-100} & {\footnotesize{}45} & {\footnotesize{}3.250} & {\footnotesize{}5.679} & {\footnotesize{}117.47} & {\footnotesize{}\citep{Puphal2017}} & {\footnotesize{}\citep{Puphal2017}}\tabularnewline
 & {\footnotesize{}-} & \textcolor{orange}{\footnotesize{}Eu}{\footnotesize{}Cu$_{3}$(OH)$_{6}$Cl$_{3}$} & \textcolor{orange}{\footnotesize{}94.7} & {\footnotesize{}P$\bar{3}$m1} & {\footnotesize{}K} & {\footnotesize{}15} & {\footnotesize{}-400} & {\footnotesize{}119.3} & {\footnotesize{}3.418} & {\footnotesize{}5.630} & {\footnotesize{}119.34} & {\footnotesize{}this paper} & {\footnotesize{}this paper}\tabularnewline
 & {\footnotesize{}-} & \textit{\textcolor{cyan}{\footnotesize{}Cd}}\textit{\footnotesize{}Cu$_{3}$(OH)$_{6}$}\textit{\textcolor{violet}{\footnotesize{}(NO$_{3}$)$_{2}$}} & \textit{\textcolor{cyan}{\footnotesize{}95, }}\textit{\textcolor{violet}{\footnotesize{}179}} & \textit{\footnotesize{}P$\bar{3}$m1} & \textit{\footnotesize{}K} & \textit{\footnotesize{}4} & \textit{\footnotesize{}45} & \textit{\footnotesize{}11} & \textit{\footnotesize{}3.261} & \textit{\footnotesize{}7.012} & \textit{\footnotesize{}106.43} & \textit{\footnotesize{}\citep{claringbullite,Cadmiumnitrat}} & {\footnotesize{}\citep{Cadmiumnitrat}}\tabularnewline
 & {\footnotesize{}-} & \textit{\footnotesize{}SmCu$_{3}$(OH)$_{6}$Cl$_{3}$} & \textit{\footnotesize{}95.8} & \textit{\footnotesize{}P$\bar{3}$m1} & \textit{\footnotesize{}T} & \textit{\footnotesize{}18} & \textit{\footnotesize{}-106} & \textit{\footnotesize{}5} & \textit{\footnotesize{}3.432} & \textit{\footnotesize{}5.639} & \textit{\footnotesize{}120.1} & \textit{\footnotesize{}\citep{Sun(2017)}} & {\footnotesize{}-}\tabularnewline
 & {\footnotesize{}-} & \textit{\textcolor{yellow}{\footnotesize{}Nd}}\textit{\footnotesize{}Cu$_{3}$(OH)$_{6}$Cl$_{3}$} & \textit{\textcolor{yellow}{\footnotesize{}98.3}} & \textit{\footnotesize{}P$\bar{3}$m1} & \textit{\footnotesize{}T} & \textit{\footnotesize{}20} & \textit{\footnotesize{}-345} & \textit{\footnotesize{}19} & \textit{\footnotesize{}3.411} & \textit{\footnotesize{}5.625} & \textit{\footnotesize{}119.30} & \textit{\footnotesize{}\citep{Sun(2017)}} & {\footnotesize{}-}\tabularnewline
 & {\footnotesize{}centennialite} & \textcolor{olive}{\footnotesize{}Ca}{\footnotesize{}Cu$_{3}$(OH)$_{6}$Cl$_{2}$} & \textcolor{olive}{\footnotesize{}100} & {\footnotesize{}P$\bar{3}$m1} & {\footnotesize{}K} & {\footnotesize{}5} & {\footnotesize{}-56} & {\footnotesize{}11} & {\footnotesize{}3.324} & {\footnotesize{}5.760} & {\footnotesize{}114.09} & {\footnotesize{}\citep{centennialite}} & {\footnotesize{}\citep{centennialite crys}}\tabularnewline
\hline 
\multirow{2}{*}{\begin{turn}{90}
{\footnotesize{}$x=2$}
\end{turn}} & {\footnotesize{}-} & \textcolor{brown}{\footnotesize{}Zn}{\footnotesize{}Cu(OH)$_{3}$Cl} & \textcolor{brown}{\footnotesize{}73} & {\footnotesize{}P2$_{1}$/m} & {\footnotesize{}1D} & {\footnotesize{}?} & {\footnotesize{}?} & {\footnotesize{}?} & {\footnotesize{}3.195} & {\footnotesize{}5.688} & {\footnotesize{}107.70} & {\footnotesize{}\citep{Znx=00003D2}} & {\footnotesize{}-}\tabularnewline
 & \textit{\footnotesize{}iyoite} & \textit{\textcolor{gray}{\footnotesize{}Mn}}\textit{\footnotesize{}Cu(OH)$_{3}$Cl} & \textit{\textcolor{gray}{\footnotesize{}83}} & \textit{\footnotesize{}P2$_{1}$/m} & \textit{\footnotesize{}1D} & \textit{\footnotesize{}29} & \textit{\footnotesize{}-80} & \textit{\footnotesize{}3} & \textit{\footnotesize{}3.302} & \textit{\footnotesize{}5.721} & \textit{\footnotesize{}114.12} & \textit{\footnotesize{}\citep{misakiite}} & {\footnotesize{}-}\tabularnewline
\hline 
\multirow{2}{*}{\begin{turn}{90}
{\footnotesize{}trimer}
\end{turn}} & {\footnotesize{}-} & \textcolor{black}{\footnotesize{}Sr}{\footnotesize{}Cu(OH)$_{3}$Cl} & {\footnotesize{}117} & {\footnotesize{}Pmn2$_{1}$} & {\footnotesize{}Trimer} & {\footnotesize{}-} & {\footnotesize{}-135} & {\footnotesize{}$\infty$} & {\footnotesize{}3.395} & {\footnotesize{}6.972} & {\footnotesize{}119.40} & {\footnotesize{}\citep{SrCu(OH)3Cl}} & {\footnotesize{}-}\tabularnewline
 & {\footnotesize{}-} & \textcolor{brown}{\footnotesize{}Zn}{\footnotesize{}Cu$_{3}$(OH)$_{6}$}\textcolor{purple}{\footnotesize{}SO$_{4}$} & \textcolor{brown}{\footnotesize{}73,}\textcolor{purple}{\footnotesize{}
258} & {\footnotesize{}P2$_{1}$/a} & {\footnotesize{}Trimer} & {\footnotesize{}-} & {\footnotesize{}-79} & {\footnotesize{}$\infty$} & {\footnotesize{}2.781} & {\footnotesize{}4.441} & {\footnotesize{}91.86} & {\footnotesize{}\citep{namuwite}} & {\footnotesize{}-}\tabularnewline
\hline 
\multirow{2}{*}{\begin{turn}{90}
{\footnotesize{}$x=4$}
\end{turn}} & {\footnotesize{}``atacamite''} & \textit{\footnotesize{}(Ni}{\footnotesize{}/}\textcolor{blue}{\footnotesize{}Mg}{\footnotesize{}/}\textit{\textcolor{gray}{\footnotesize{}Mn}}{\footnotesize{})$_{2}$(OH)$_{3}$Cl} & \textit{\footnotesize{}69/ }\textcolor{blue}{\footnotesize{}72}\textit{\footnotesize{}/
}\textit{\textcolor{gray}{\footnotesize{}83}} & {\footnotesize{}Pnma} & {\footnotesize{}P} & {\footnotesize{}5/}\textcolor{darkgray}{\footnotesize{}3.4} & \textit{\footnotesize{}-100}{\footnotesize{}/}\textit{\textcolor{darkgray}{\footnotesize{}-57.8}} & \textit{\footnotesize{}20}{\footnotesize{}/}\textit{\textcolor{darkgray}{\footnotesize{}17}} & {\footnotesize{}3.359} & {\footnotesize{}5.190} & {\footnotesize{}107.7} & {\footnotesize{}\citep{M2(OH)3Cl,Ni2(OH)3Cl,Mn2(OH)3Cl}} & {\footnotesize{}-}\tabularnewline
 & {\footnotesize{}``kapellasite''} & {\footnotesize{}(Ni/Co/}\textcolor{blue}{\footnotesize{}Mg}{\footnotesize{}/}\textcolor{violet}{\footnotesize{}Fe}{\footnotesize{}/}\textcolor{gray}{\footnotesize{}Mn}{\footnotesize{})$_{2}$(OH)$_{3}$Cl} & \textit{\footnotesize{}69/ 74.5/ }\textit{\textcolor{violet}{\footnotesize{}75}}\textit{\textcolor{black}{\footnotesize{}/}}\textit{\footnotesize{}
}\textcolor{blue}{\footnotesize{}72}\textit{\footnotesize{}/ }\textit{\textcolor{gray}{\footnotesize{}83}} & {\footnotesize{}P$\bar{3}$m1} & {\footnotesize{}T} & {\footnotesize{}?} & {\footnotesize{}?} & {\footnotesize{}?} & {\footnotesize{}?} & {\footnotesize{}?} & {\footnotesize{}?} & {\footnotesize{}\citep{M2(OH)3Cl}} & -\tabularnewline
\end{tabular} 
\end{spacing}

\caption{\label{tab:List}List of the atacamite family divided into four classes
of structures each sorted by the size of the substitution ions. The
table includes the mineral name if naturally existing, the chemical
formula, the 6 fould coordinated radius of the substituted ion, the
structural space group, the lattice type of the magnetic ion (K: kagome,
P: pyrochlore, T: triangular and 1D: chains), the magnetic transition
temperature, the Curie-Weiss temperature with the resulting frustration
as well as the Cu arrangements. Column 10-12 describe the structure
with first the relevant magnetic ion distance, then the distance between
the sought kagome layers (ignoring the position in between) and finally
the highest Cu-O-Cu bonding angle. In the next-to-last column first
the reference for the magnetic properties is given, then the one of
the structure and in the last column a reference for single crystal
growth. If the substitution ion itself is magnetic the whole line
is written in italic.}
\end{sidewaystable}

\twocolumngrid

\section{Experimental Detail}

\textcolor{black}{For the single crystal growth a hydrothermal Parr
4625 autoclave with a 575 ml filling capacity operated by a Parr 4842
power supply including a 982 Watlow controller was used.}

\textcolor{black}{The energy dispersive X-ray spectra (EDS) were recorded
with an }AMETEK EDAX Quanta 400\textcolor{black}{{} detector in a }Zeiss
DSM 940A\textcolor{black}{{} scanning electron microscope (SEM). A layer
of carbon was sputtered on the isolating single crystals using a Balzers
Union FL-9496.}

\textcolor{black}{X-ray powder diffraction data were collected on
a Bruker D8 Focus using a Cu X-ray generator and the Rietveld refinement
of the X-ray data was done using the fullprof suite \citep{Rodriguez(1993)}.
For the single crystal structure determination the data were collected
at 173~K on a STOE IPDS II two-circle diffractometer with a Genix
Microfocus tube with mirror optics using Mo K$_{\alpha}$ radiation
($\lambda=0.71073\thinspace\textrm{\AA}$). The data were scaled using
the frame scaling procedure in the X-AREA program system \citep{Stoe(2002)}.
The structure was solved by direct methods using the program SHELXS
\citep{Sheldrick(2008)} and refined against $F^{2}$ with full-matrix
least-squares techniques using the program SHELXL-97 \citep{Sheldrick(2008)}.
The H atoms bonded to O were found in a difference map and was isotropically
refined with the O-H distances restrained to 0.84(1)\,Å.}

\textcolor{black}{The specific-heat and magnetic measurements were
collected with the standard options of a Physical Property Measurement
System from Quantum Design in a temperature range of 1.8 to 350\,K.}

\section{$\text{EuCu}$$_{3}$$\text{(OH)}$$_{6}$$\text{Cl}$$_{2}$}

\subsection{Synthesis}

\textcolor{black}{Single crystals of }EuCu$_{3}$(OH)$_{6}$Cl$_{3}$\textcolor{black}{{}
were prepared in the Parr autoclave with a temperature profile adapted
from the optimised one for Y$_{3}$Cu$_{9}$(OH)$_{19}$Cl$_{8}$
\citep{Puphal2017}. For the crystallization, we prepared duran glass
ampoules as follows: The ampoules were loaded with 0.6513~g CuO and
2~g EuCl$_{3}\cdot6$H$_{2}$O, solved in 5~ml distilled water and
then sealed at air. They were placed in the autoclave, which was filled
with distilled water to ensure the same pressure as in the ampoules.
The autoclave was heated up to 270\textdegree C in four hours and
subsequently cooled down to 170\textdegree C with 1 K/h, followed
by a fast cooling to room temperature. Afterwards, the ampoules were
opened and the content was filtered with distilled water. The ampoules
contained a few larger single crystals, some smaller ones and a pellet.
The crystals have blue colour and a hexagonal shape with typical sizes
up to 1 x 1 x 0.25 mm$^{3}$ (see figure \ref{Eu structure} c). Similar
to Y$_{3}$Cu$_{9}$(OH)$_{19}$Cl$_{8}$ \citep{Puphal2017} and
haydeeite, the crystal habitus gives information about all crystallographic
axis, with the a and b axis being the corners of the hexagon and the
c axis perpendicular to the surface of the platelets. Attempts with
lower and higher EuCl$_{3}\cdot6$H$_{2}$O content in the solutions
were also successfull but lead to smaller single crystals. However,
the Eu content does not change with varying salt amount, which was
deduced from EDS. }

\textcolor{black}{Furthermore, a growth in an external temperature
gradient }similar to \citep{Han(2011)} led to large single crystals
of up to \textcolor{black}{2 x 2 x 0.75 mm$^{3}$} (shown in figure
\ref{Eu structure} b). We used\textcolor{black}{{} }2~g\textcolor{black}{{}
pre-reacted} EuCu$_{3}$(OH)$_{6}$Cl$_{3}$\textcolor{black}{{} in
a solution of }1~g \textcolor{black}{EuCl$_{3}\cdot6$H$_{2}$O}
with 5~ml H$_{2}$O sealed in a thick walled quartz ampoule of 15~cm
length. Then we placed this ampoule into an external gradient of 2\textdegree C/cm,
with 250\textdegree C at the hot end and 220\textdegree C at the cold
end. After several weeks the whole powder was recrystallized. 

\subsection{EDS Analysis}

\begin{figure}[H]
\noindent \begin{centering}
\includegraphics[width=1\columnwidth]{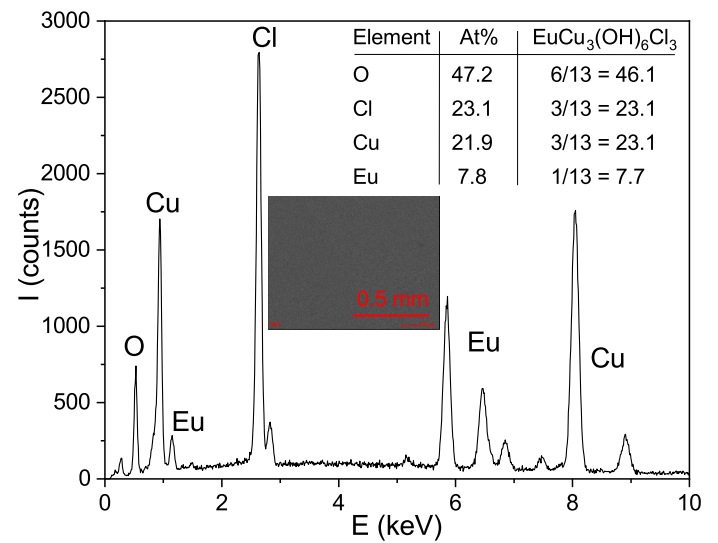}
\par\end{centering}
\caption{\textcolor{black}{\label{EDX}EDS analysis on a polished single crystal
of }EuCu$_{3}$(OH)$_{6}$Cl$_{3}$\textcolor{black}{. In a table
the measured at\% are given compared to the stoichiometric values,
not including the H atoms. The inset shows a SEM image, which would
have shown impurity phases as grey-scaled contrast, proving the homogeneity
of the crystal.}}
\end{figure}

We measured EDS in the SEM on several crystals of different batches
both of polished and untreated single crystals. In figure \ref{EDX}
an EDS spectra is shown with the SEM image on the inset showing a
clean surface with no impurity phases. The resulting at\% are given
in the figure and are in overall agreement with stoichiometric values
of EuCu$_{3}$(OH)$_{6}$Cl$_{3}$. We measured several points on
many crystals of different batches and did not observe any variance
regarding the Eu-Cu ratio. This is typical for the P-3m1 structure,\textcolor{black}{{}
which has no intermediate substitutions and no paratacamite like crossover
structure. E.g. kapellasite and haydeeite \citep{Colman(2008),Colman(2011)}
stabilizes a certain substitution amount above one.} By directly comparing
only the Cu/Eu ratio we found the Eu amount, assuming Eu$_{x}$Cu$_{4-x}$(OH)$_{6}$Cl$_{3}$,
to be not increased with an average of $x_{Eu}=1.01\pm0.01$. 

\subsection{Structure}

\begin{figure}[H]
\noindent \begin{centering}
\includegraphics[width=1\columnwidth]{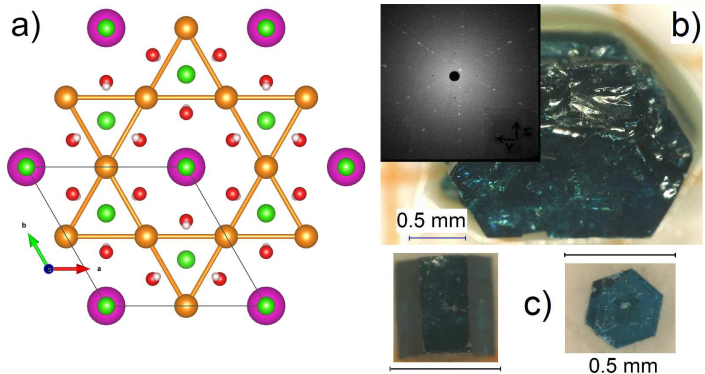}
\par\end{centering}
\caption{\textcolor{black}{\label{Eu structure}}a) Image of the P-3m1 structure
viewed onto the ab plane. There, the orange star like kagome arrangement
of the Cu atoms is apparent. The pink europium atoms are positioned
in the center of the star in the kagome plane typical for the kapellasite
P-3m1 structure. b) Laue-Image and a picture of a large EuCu$_{3}$(OH)$_{6}$Cl$_{3}$
crystal with a coaligned twin obtained from an external gradient growth.
c) Image of two examplous untreated blue EuCu$_{3}$(OH)$_{6}$Cl$_{3}$
single crystals \textcolor{black}{from a growth in the autoclave}
showing nice hexagonal shape.}
\end{figure}

EuCu$_{3}$(OH)$_{6}$Cl$_{3}$ crystallizes in the kapellasite type
structure P-3m1 with an additional Cl position just like Y, Nd, Sm
and Gd \citep{Puphal2017,Sun(2016),Sun(2017)}. The structure was
obtained by single crystal diffraction and confirmed with powder diffraction.
It is shown in figure \ref{Eu structure} a) viewed along the c-axis
revealing the kagome star of the orange Cu atoms. The distances are
$d_{Cu1-Cu1}=3.4181(4)\textrm{\thinspaceÅ}$ of the Cu atoms in the
kagome plane and $d_{Cu1-Cu2}=5.6301(13)\thinspace\text{Å}$ between
the AA stacked planes. The angle for the dominant superexchange is
$\angle$(Cu1-O1-Cu1) = 119.34(16)\textdegree . The resulting blue
crystals shown in figure \ref{Eu structure} b, c) have a hexagonal
shape directly revealing the structure enabling an easy orientation.
The atomic positions are given in table \ref{tab:Structural-table-of}:

\begin{table}[h]
{\footnotesize{}}%
\begin{tabular}{l|lllllll}
{\footnotesize{}position} & {\footnotesize{}x/a} & {\footnotesize{}y/b} & {\footnotesize{}z/c} & {\footnotesize{}Occ.} & {\footnotesize{}U$_{iso}$} & {\footnotesize{}Site} & {\footnotesize{}Sym.}\tabularnewline
\hline 
{\footnotesize{}Eu1} & {\footnotesize{}0} & {\footnotesize{}0} & {\footnotesize{}0.5} & {\footnotesize{}1} & {\footnotesize{}0.0131(3)} & {\footnotesize{}1b} & {\footnotesize{}-3m.}\tabularnewline
{\footnotesize{}Cu1} & {\footnotesize{}0.5} & {\footnotesize{}0} & {\footnotesize{}0.5} & {\footnotesize{}1} & {\footnotesize{}0.0123(4)} & {\footnotesize{}3f} & {\footnotesize{}.2/m.}\tabularnewline
{\footnotesize{}Cl1} & {\footnotesize{}0.666667} & {\footnotesize{}0.333333} & {\footnotesize{}0.1359(4)} & {\footnotesize{}1} & {\footnotesize{}0.0190(6)} & {\footnotesize{}2d} & {\footnotesize{}3m.}\tabularnewline
{\footnotesize{}Cl2} & {\footnotesize{}0} & {\footnotesize{}0} & {\footnotesize{}0} & {\footnotesize{}1} & {\footnotesize{}0.0176(7)} & {\footnotesize{}1a} & {\footnotesize{}-3m.}\tabularnewline
{\footnotesize{}O1} & {\footnotesize{}0.8066(4)} & {\footnotesize{}0.1934(4)} & {\footnotesize{}0.6318(7)} & {\footnotesize{}1} & {\footnotesize{}0.0122(8)} & {\footnotesize{}6i} & {\footnotesize{}.m.}\tabularnewline
{\footnotesize{}H1} & {\footnotesize{}0.790(8)} & {\footnotesize{}0.210(8)} & {\footnotesize{}0.776(4)} & {\footnotesize{}1} & {\footnotesize{}-} & {\footnotesize{}6i} & {\footnotesize{}.m.}\tabularnewline
\end{tabular}{\footnotesize \par}

\caption{\label{tab:Structural-table-of}Structural table obtained by refining
single crystal diffraction data on a EuCu$_{3}$(OH)$_{6}$Cl$_{3}$
single crystal measured at 173\,K. A trigonal P-3m1 (\#164) structure
with a unit cell of $a=b=6.8363(14)\textrm{\thinspaceÅ}$ and $c=5.6301(13)\thinspace\textrm{Å}$,
$\alpha=\beta=90\text{\textdegree}$, $\gamma=120\text{\textdegree}$
was found, where the occupation was fixed to 1.}

\end{table}

\subsection{Magnetic susceptibility}

\begin{figure}[h]
\noindent \begin{centering}
\includegraphics[width=1\columnwidth]{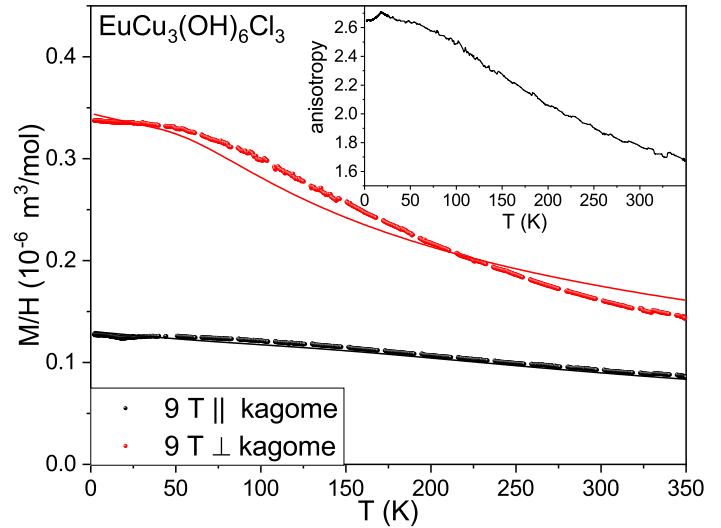}
\par\end{centering}
\caption{\textcolor{black}{\label{fig:vanvleck}The image shows the magnetic
data on a stack of four single crystals with a mass of 2\,mg of the
temperature dependant susceptibility data at 9 T with the field aligned
(black dots) and perpendicular to the kagome plane (red dots). The
line is obtained by a fit using both Van Vleck and Curie-Weiss contribution.
On the inset the temperature dependant anisotropy of perpendicular/
parallel susceptiblity is shown.}}
\end{figure}

In this section, we present a thorough magnetic analysis on EuCu$_{3}$(OH)$_{6}$Cl$_{3}$
single crystals. To obtain larger masses for a reduced noise ratio,
four single crystals were aligned with either the magnetic field parallel
(\textbar{}\textbar{}) or perpendicular ($\perp$) to the kagome plane.
We start with the analysis of the susceptibility at 9\,T for temperatures
between 2 and 350\,K, shown in figure \ref{fig:vanvleck}. In this
temperature regime Eu$^{3+}$ ions cause pronounced Van Vleck paramagnetism,
which gives a separate magnetic contribution in addition to the magnetism
from the Cu$^{2+}$. The susceptibility in this temperature range
reveals a pronounced anisotropy, $\chi_{||}/\chi_{\bot}$ from 1.7
at 350\,K to 2.5 at 50\,K plotted in the inset of figure \ref{fig:vanvleck}.
This anisotropy is most likely attributed to the Van Vleck contribution,
similar as it was observed in Ref. \citep{Tovar(1989),Yamaguchi(1988)}.
There, the anisotropy arises from a splitting of the $^{1}$F$_{1}$
level due to a crystal electric field. In literature this anisotropy
was analyzed for a tetragonal symmetry of the Eu$^{3+}$, while in
EuCu$_{3}$(OH)$_{6}$Cl$_{3}$ the europium atoms are located on
a site with hexagonal symmetry. We fitted the experimental data in
the full range from 1.8\,K to 350\,K at a high field of 9\,T in
figure \ref{fig:vanvleck} with the sum of a general Van Vleck part
(see Ref. \citep{VanVleck}) and the Curie-Weiss contribution from
three Cu$^{2+}$ ions following the formula

{\small{}
\[
\chi_{mol}=3\frac{C}{T+\Theta_{W}}+\frac{N_{A}\mu_{0}\mu_{B}^{2}}{3\lambda}\left(\frac{24+(13.5\frac{\lambda}{k_{B}T}-1.5)\text{e}^{-\lambda/(k_{B}T)}+\ldots}{1+3\text{e}^{-\lambda/(k_{B}T)}+\ldots}\right)
\]
}{\small \par}

to account for the different magnetic contributions. We note however
that the two contributions are hard to seperate and the obtained values
are only an estimate, since a nonmagnetic reference e.g. EuZn$_{3}$(OH)$_{6}$Cl$_{3}$
would be necessary to accurately define the Eu$^{3+}$ Van Vleck contributions.
Assuming a generally similar behavior as for the tetragonal symmetry,
we get a spin orbit value of roughly $\lambda=(2\lambda_{\Vert}+\lambda_{\bot})/3=(2\cdot210+586)/3\thinspace\text{K}=335$\,K
comparable to Eu$_{2}$CuO$_{4}$ with $\lambda=315$\,cm$^{-1}\approx300$\,K
\citep{Tovar(1989)}. The Curie-Weiss temperatures was calculated
in this global fit yielding roughly $\Theta_{W}\approx-400$\,K.\\
From a $\sqrt{\chi T}$ plot (not shown) the whole material reveals
increased effective moment values of $\mu_{eff}^{||}=3.25$\,$\mu_{B}$
and $\mu_{eff}^{\perp}=2.5$\,$\mu_{B}$ at 350\,K compared to the
theoretical one for Cu$^{2+}$ of 1.73\,$\mu_{B}$ as a result of
the mixture with the Van Vleck contribution of Eu$^{3+}$.

\begin{figure}[h]
\noindent \begin{centering}
\includegraphics[width=1\columnwidth]{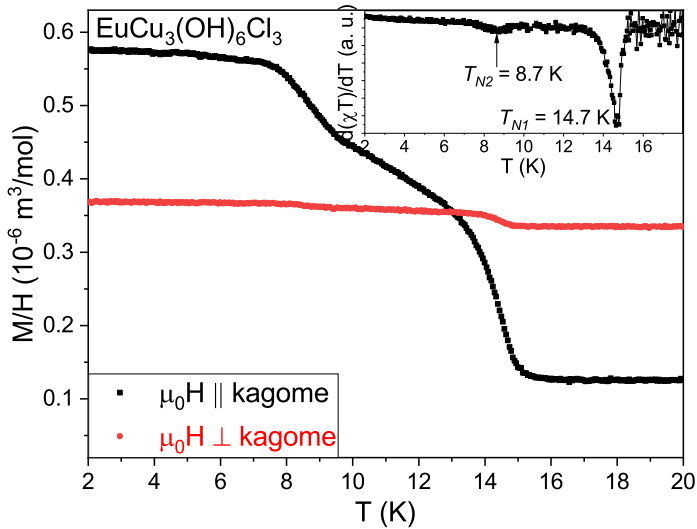}
\par\end{centering}
\caption{\textcolor{black}{\label{order}The main figure shows the low temperatue
part of the susceptibility data of EuCu$_{3}$(OH)$_{6}$Cl$_{3}$
field cooled at 0.1\,T both measured with the field aligned (red)
and perpendicular to the kagome plane (black). Inset: Derivative of
$\chi\cdot T$ at a tiny field of 50 Oe. }}
\end{figure}

In figure \ref{order} we present the low temperature behaviour of
the susceptibility of EuCu$_{3}$(OH)$_{6}$Cl$_{3}$, which is dominated
by the Cu magnetism. Two magnetic transitions are apparent at $T_{N_{1}}=14.7$\,K
and $T_{N_{2}}=8.7$\,K for both field directions. The transition
temperatures can best be determined by the derivate $d(\chi T)/dT$
at very small field, shown in the inset of figure \ref{order}. We
observed a small sample dependant shift of these magnetic transition
of $\Delta T=\pm1$\,K for the higher and $\Delta T=\pm2$\,K for
the low transition, when comparing measurements on different single
crystals. The Weiss temperatures hint to a dominant antiferromagnetic
exchange between the Cu ions of 400\,K, but the order appears only
below 15\,K. Therefore, EuCu$_{3}$(OH)$_{6}$Cl$_{3}$ is a strongly
frustrated system with a frustration coefficient $f=|\vartheta_{W}|/T_{C}\approx30$.

\begin{figure}[h]
\noindent \begin{centering}
\includegraphics[width=1\columnwidth]{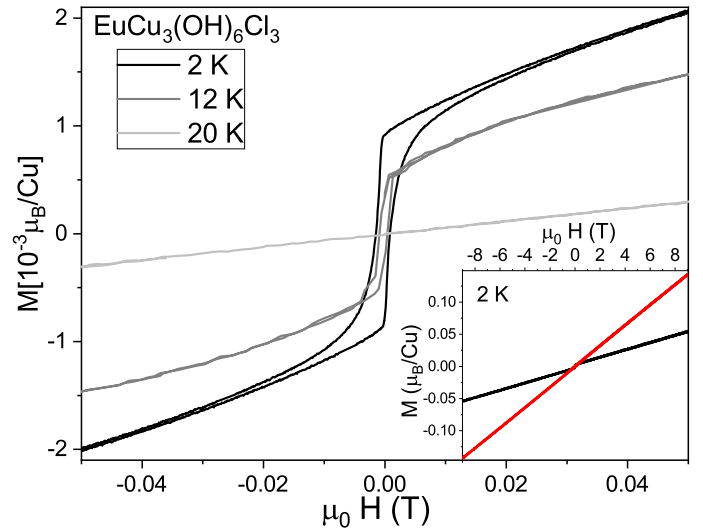}
\par\end{centering}
\caption{\textcolor{black}{\label{hysteresis}Excerpt of the M(H) curves measured
at 2\,K (black), 12\,K (dark grey) and 20\,K (light grey) with
a field applied parallel to the kagome plane. The inset shows the
full range for a measurement at 2\,K for both parallel (black) and
perpendicular fields (red).}}
\end{figure}

Insight into the nature of the magnetic transition can be obtained
from the magnetization curves between 2 and 20\,K shown in figure
\ref{hysteresis}. A small but well defined hysteresis can be seen
for magnetic field within the kagome plane. However, the spontaneous
moment below $T$$_{N_{1}}$ is only $0.5\cdot10{}^{-3}$\,$\mu_{B}$
per Cu and $1\cdot10{}^{-3}$\,$\mu_{B}$ below $T$$_{N_{2}}$.
This shows, that the magnetic order has a tiny in-plane ferromagnetic
component. For fields perpendicular to the kagome layers nearly no
hysteresis was observed. At higher magnetic fields, both directions
present a linear $M(H)$ dependence, with a higher slope for the direction
perpendicular to the kagome plane. At 9\,T, M reaches only about
0.05 and 0.14 \textmu B/Cu for the two field directions, which is
well below the saturation magnetization of 1\,$\mu_{B}$ per Cu$^{2+}$
ion. Furthermore, also the Van Vleck contribution from Eu might contribute
to the magnetization at higher fields, but is difficult to entangle
from the Cu magnetism. For that purpose a reference compound with
Zn instead of Cu would be necessary, which however, is not known to
exist so far. 

\subsection{Specific heat}

\begin{figure}[h]
\noindent \begin{centering}
\includegraphics[width=1\columnwidth]{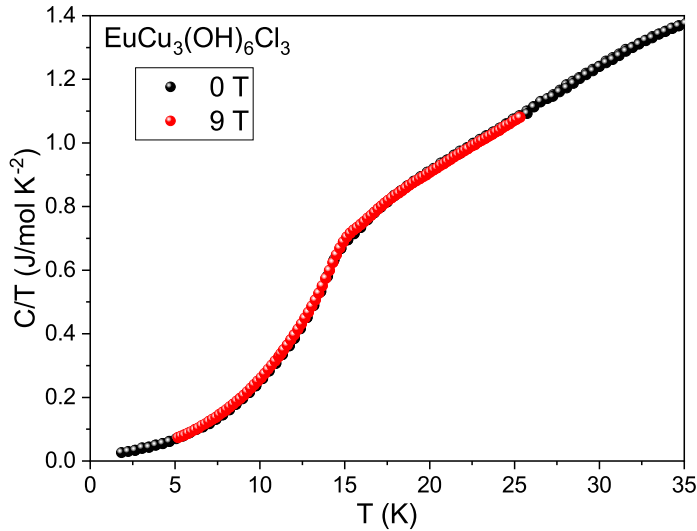}
\par\end{centering}
\caption{\textcolor{black}{\label{spec heat Ga}Specific heat measurements
of EuCu$_{3}$(OH)$_{6}$Cl$_{3}$ in 0 and 9\,T perpendicular to
the kagome plane measured from 1.8 to 39\,K on a single crystal of
8.61\,mg. A broad anomaly is visible at the magnetic ordering temperature.}}
\end{figure}

The observed magnetic transitions in the magnetic measurements were
also analyzed on several crystals by specific heat. This is essential
to exclude contributions from small foreigen phases to be the origin
of the observed magnetic signals. We show a measurement between 1.8\,K
and 39\,K on a single crystal of 8.61\,mg. The general specific
heat curve divided by temperature is shown in figure \ref{spec heat Ga}
for a field of 0 and 9\,T. We roughly estimated this phonon part
in a temperature range 20 - 32\,K from a linear fit of C/T vs T$^{2}$
plot. With $\beta=0.625(6)$\,mJ/mol we get an estimated debye temperature
of $\Theta_{D}\approx390$\,K slightly above the one of Y$_{3}$Cu$_{9}$(OH)$_{19}$Cl$_{8}$
\citep{Puphal2017}. We notice that this estimate is very rough and
only used to show an order of magnitude for the phon contribution.
The reason is apparent from figure \ref{spec heat Ga}, because also
at 30\,K a clear $T^{3}$ dependance of the specific heat is not
obtained, due to magnetic contributions from the fluctuating moments,
hindering a more accurate fit of the phonon contribution. At 15\,K
we observe a broad peak in the specific heat data, proving that the
transition observed in the magnetic measurements is an intrinsic property
of EuCu$_{3}$(OH)$_{6}$Cl$_{3}$. This transition shows no field
dependance for fields up to 9\,T. This is in line with the M(H) curve
up to 9\,T, where no saturation was reached, showing that the field
scale of 9\,T is far below the energy scale of the dominant magnetic
coupling. The entropy connected with the small peak is well below
the full ordered moment of Cu, in agreement with the large frustration
coefficient found in the magnetic measurements. 

\section{Z$\text{n}$$_{x}$C$\text{u}$$_{4-x}$(OH)$_{6}$(NO$_{3}$)$_{2}$}

\subsection{Synthesis}

Powder samples of\textcolor{black}{{} }ZnCu(OH)$_{6}$(NO$_{3}$)$_{2}$
were prepared by a solid state reaction of\textcolor{black}{{} 0.337g
ZnO with 1g Cu(NO$_{3}$)$_{2}-6$ H$_{2}$O and 0.6585g CuO sealed
in a Parr 4749 acid digestion vessel heated to 220\textdegree C for
two days then rapidly cooled to room temperature. }

\begin{figure}[H]
\noindent \begin{centering}
\includegraphics[width=1\columnwidth]{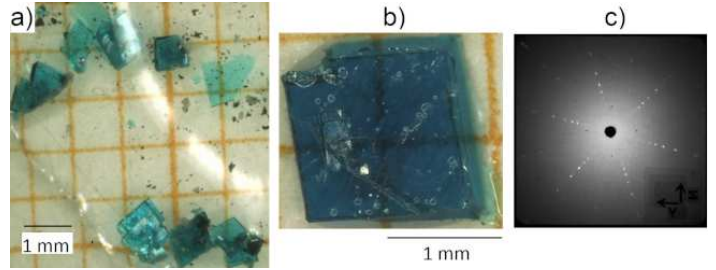}
\par\end{centering}
\caption{\textcolor{black}{\label{zn}}a), b) Picture of ZnCu$_{3}$(OH)$_{6}$(NO$_{3}$)$_{2}$
crystals from a hydrothermal growth by slow cooling. c) Laue Image
of the crystal depicted in b).}
\end{figure}

\textcolor{black}{Single crystals of }Zn$_{x}$Cu$_{4-x}$(OH)$_{6}$(NO$_{3}$)$_{2}$\textcolor{black}{{}
were prepared in the Parr autoclave with a temperature profile adapted
from the optimised one from Y$_{3}$Cu$_{9}$(OH)$_{19}$Cl$_{8}$
\citep{Puphal2017}. For the crystallization, we placed duran glass
ampoules filled with the solution in the autoclave and filled it with
distilled water to ensure the same pressure as in the ampoules. The
ampoules were loaded with 0.4\,g CuO and 2-4\,g Zn(NO$_{3}$)$_{2}$$\cdot6$
H$_{2}$O (see table \ref{tab:Results-of-substitution}) solved in
3~ml distilled water and then sealed at air. The autoclave was heated
up to 270\textdegree C in four hours and subsequently cooled down
to 140\textdegree C with 1.4 K/h, followed by a fast cooling to room
temperature. Afterwards, the ampoules were opened and the content
was filtered with distilled water. The ampoules contained a few larger
single crystals, some smaller ones and a pellet. The crystals have
blue colour and a square shape with typical sizes up to 1 x 1 x 0.25
mm$^{3}$, where some are depicted in figure \ref{zn}.}

\subsection{EDS Analysis}

We measured EDS spectra on several crystals of different batches with
different molar amounts of Zn(NO$_{3}$)$_{2}$.\textcolor{black}{{}
For this compound evaluable intensity is only caused by Zn, Cu, and
O and we found a stoichiometry Zn$_{x}$Cu$_{4-x}$(OH)$_{6}$(NO$_{3}$)$_{2}$
with a variation of the Zn to Cu ratio, depending on the initial content
of Zn(NO$_{3}$)$_{2}$ in the solution. Some examples of the substitution
amount vs Zn(NO$_{3}$)$_{2}$ amount are given in table \ref{tab:Results-of-substitution},
which show a nearly linear increase controlled by the molar ratio,
similar as for herbertsmithite, where a strong excess of Zn$^{2+}$
ions is necessary to reach substitutions of $x=1$.}

\textcolor{black}{}
\begin{table}[h]
\textcolor{black}{}%
\begin{tabular}{c|cccc}
\textcolor{black}{Zn(NO$_{3}$)$_{2}$$\cdot6$ H$_{2}$O {[}g{]}} & 2 & 2.5 & 3 & 4\tabularnewline
\hline 
\textless{}x$_{Zn}$\textgreater{} & \textcolor{black}{0.32(9)} & \textcolor{black}{0.50(12)} & \textcolor{black}{0.61(9)} & \textcolor{black}{0.86(4)}\tabularnewline
\end{tabular}

\textcolor{black}{\caption{\label{tab:Results-of-substitution}Results of substitution amounts
obtained by EDS analysis on single crystals of different batches prepared
with different molar amounts of\textcolor{black}{{} Zn(NO$_{3}$)$_{2}$.
The error for $x_{Zn}$ is a statistical one, obtained by measuring
several spots on several crystals from the same batch. }}
}
\end{table}

\subsection{Structure}

\begin{figure}[H]
\noindent \begin{centering}
\includegraphics[width=1\columnwidth]{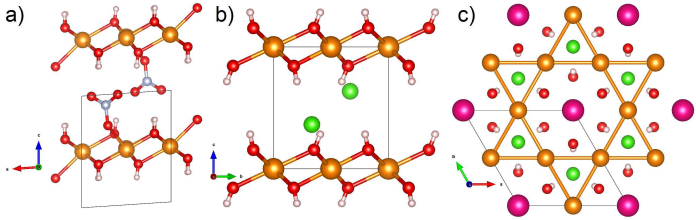}
\par\end{centering}
\caption{\textcolor{black}{\label{zn-struct}}a) Picture of the Zn$_{x}$Cu$_{4-x}$(OH)$_{6}$(NO$_{3}$)$_{2}$
structure viewed along the b-axis. b) Picture of the haydeeite structure
viewed along the a-axis. The colorcode is: A = Zn, Cu, Mg (orange),
O (red), Cl (green), N (grey) and H (white). c) Picture of the haydeeite
structure viewed along the c-axis now with Mg in pink.}
\end{figure}

From table \ref{tab:List} it can be seen, that presently two variants
in that family with the NO$_{3}$ anion were reported in literature.
Rouaite, Cu$_{4}$(OH)$_{6}$(NO3)$_{2}$ with space group P2$_{1}$
and CdCu$_{3}$(OH)$_{6}$(NO$_{3}$)$_{2}$ with space group P$\bar{3}$m1
similar to haydeeite. It is difficult to differentiate Zn and Cu from
X-rays, thus we found the structure of rouaite as the supergroup P2$_{1}$
(\#4) to be the correct cell by taking only Cu atoms on the Zn-Cu
positions, A$_{4}$(OH)$_{6}$(NO$_{3}$)$_{2}$ with A = Zn, Cu.
The structure generally describes the matrix and was obtained by single
crystal diffraction. To envision the similarities in figure \ref{zn-struct}
the structure is shown along the b-axis compared to the a-axis of
haydeiite with Mg changed to Cu as it is chosen for the P2$_{1}$
structure. From the image one can easily see the similar arrangement
of the A ions and the comparable positions of Cl and NO$_{3}$. Thus
a similar structure is expected for \textcolor{black}{Zn$_{x}$Cu$_{4-x}$(OH)$_{6}$(NO$_{3}$)$_{2}$,
as depicted in figure \ref{zn-struct} c).} To differentiate between
Zn and Cu resonant X-rays are necessary to find the correct subgroup.

The distances are $d_{A-A}=3.0648(4)\textrm{\thinspaceÅ}$ of the
Cu atoms in the kagome plane and $d_{A-A}=6.9271(9)\thinspace\text{Å}$
between the AA stacked planes. The angle for the dominant superexchange
is $\angle$(Cu1-O1-Cu1) = 104.07(12)\textdegree . The atomic positions
are given in table \ref{tab:Structural-table-of-Zn}:

\begin{table}[h]
\begin{centering}
\begin{tabular}{l|lllllll}
 & x/a & y/b & z/c & Occ. & U$_{iso}$ & Site & Sym.\tabularnewline
\hline 
A1 & 0.49908 & 0.50073 & 0.49403 & 1 & 0.007 & 2a & 1\tabularnewline
A2 & -0.01041 & 0.25265 & 0.50224 & 1 & 0.007 & 2a & 1\tabularnewline
O1 & 0.3664 & 0.7556 & 0.3577 & 1 & 0.009 & 2a & 1\tabularnewline
H1 & 0.384 & 0.773 & 0.240 & 1 & 0.014 & 2a & 1\tabularnewline
O2 & 0.8141 & 0.5099 & 0.3755 & 1 & 0.008 & 2a & 1\tabularnewline
H2 & 0.819 & 0.513 & 0.255 & 1 & 0.012 & 2a & 1\tabularnewline
O3 & 0.1877 & 0.4984 & 0.6212 & 1 & 0.009 & 2a & 1\tabularnewline
H3 & 0.194 & 0.483 & 0.743 & 1 & 0.013 & 2a & 1\tabularnewline
O4 & 0.2937 & 0.2596 & 0.2789 & 1 & 0.013 & 2a & 1\tabularnewline
O5 & 0.1092 & 0.1561 & 0.0086 & 1 & 0.031 & 2a & 1\tabularnewline
O6 & 0.4036 & 0.3871 & 0.0063 & 1 & 0.026 & 2a & 1\tabularnewline
N1 & 0.2680 & 0.2675 & 0.0942 & 1 & 0.015 & 2a & 1\tabularnewline
\end{tabular}
\par\end{centering}
\caption{\label{tab:Structural-table-of-Zn}Structural table obtained by refining
(occupation fixed to 1) the single crystal diffraction data on a\textcolor{black}{{}
}Zn$_{0.7}$Cu$_{3.3}$(OH)$_{6}$(NO$_{3}$)$_{2}$ single crystal
measured at 173\,K. Using the monoclinic P2$_{1}$ (\#4) structure
with a unit cell of $a=5.5704(6)\textrm{\thinspaceÅ}$, $b=6.1274(7)\textrm{\thinspaceÅ}$
and $c=6.9271(8)\thinspace\textrm{Å}$, $\alpha=\gamma=90\text{\textdegree}$,
$\beta=93.782(9)\text{\textdegree}$.}
\end{table}

\subsection{Magnetic susceptibility}

\begin{figure}[h]
\noindent \begin{centering}
\includegraphics[width=1\columnwidth]{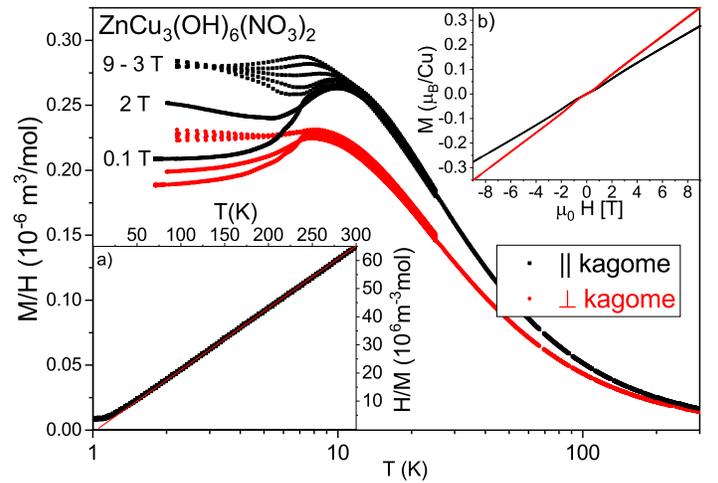}
\par\end{centering}
\caption{\textcolor{black}{\label{susz Zn}The image shows the magnetic data
on a Zn$_{0.6}$Cu$_{3.4}$(OH)$_{6}$(NO$_{3}$)$_{2}$ single crystal
with a mass of 2.23\,mg of the susceptibility in the range of 1.8
- 350\,K field cooled at 0.1 - 9\,T with the field aligned (black
dots) and perpendicular to the kagome plane (red dots). a}) Inverse
susceptibility $(M/H)$$^{-1}$ measured on a powder sample of 49.1\,mg.
b) $M(H)$ curve at 2\,K for both field directions, which lay nearly
perfectly on each other.}
\end{figure}

We performed magnetic measurements on various Zn$_{x}$Cu$_{4-x}$(OH)$_{6}$(NO$_{3}$)$_{2}$
single crystals. We found that for concentrations 0.3 \textless{}
x \textless{} 0.9 the magnetic behavior and the ordering temperature
is only slightly changed with varying x. In figure \ref{susz Zn}
we show exemplarily the measurement on a single crystal with $x=0.6$.
The overall anisotropy is very small, as can be seen from the magnetization
at 2 K up to 9T, where both curves for field parallel and perpendicular
to the kagome completely overlap. The susceptibility at 0.1\,T shows
a well-defined anomaly at $T_{N}\approx7.5$\,K indicating the onset
of antiferromagnetic order. The field dependence of the susceptibility
below $T_{N}$ shows a small anisotropy for magnetic fields below
3\,T, this is also reflected in small metamagetic transitions in
M(H) in this field range. 

From a $\sqrt{\chi T}$ plot (not shown) the effective moment values
of $\mu_{eff}^{||}=$1.62\,$\mu_{B}$ and $\mu_{eff}^{\perp}=$1.65\,$\mu_{B}$
at 350\,K can be derived. A Curie-Weiss fit of the $x=1$ powder
sample is shown in figure \ref{susz Zn} b) and a Curie-Weiss temperature
of $\Theta_{W}\approx$5\,K, which indicates contributions both from
ferromagnetic and antiferromagnetic exchange couplings.

The $M(H)$ curve above 3\,T reveals nearly perfect linear behavior
with a polarized moment of about 0.3\,$\mu_{B}$ at 9\,T, still
well below the saturated moment of 1\,$\mu_{B}$ per Cu$^{2+}$.
The overall shape of the $M(H)$ curve confirms the antiferromagnetic
nature of the ordered state. 

\subsection{Specific heat}

\begin{figure}[h]
\noindent \begin{centering}
\includegraphics[width=1\columnwidth]{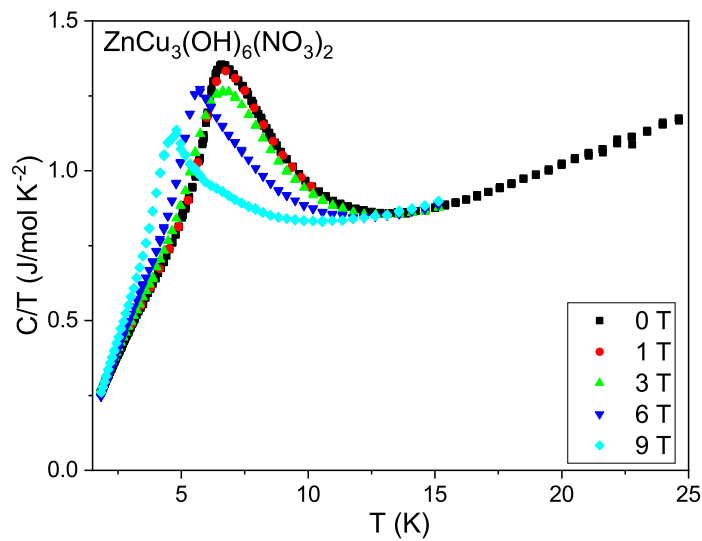}
\par\end{centering}
\caption{\textcolor{black}{\label{spec heat Zn}Specific heat measurements
in an external field perpendicular to the plane of 0, 1, 3, 6 and
9\,T from 1.8 to 25\,K on a single crystal of 5.1\,mg. A clear
anomaly is visible at the magnetic ordering temperature.}}
\end{figure}

We measured the magnetic transition on a single crystal by specific
heat. Shown is a measurement within 1.8\,K and 15\,K on a single
crystal with $x=0.6$ and a mass of 8.61\,mg. The specific heat curve
divided by temperature is shown in figure \ref{spec heat Zn} for
fields from 0 up to 9\,T. We roughly estimated the phonon part in
a temperature range of 15 - 25\,K from a linear fit of a C/T vs T$^{2}$
plot. With $\beta=0.800(6)$\,mJ/mol we get an estimated debye temperature
of $\Theta_{D}\approx387$\,K. We observe a pronounced peak at $T_{N}$,
which is slowly shifted to lower temperatures with increasing fields
and reaches 5~K at 9\,T. The ordered fraction connected to the anomaly
is much higher, compared to the EuCu$_{3}$(OH)$_{6}$Cl$_{3}$ system
and proves that \textcolor{black}{ZnCu(OH)$_{6}$(NO$_{3}$)$_{2}$}
is only slightly geometrically frustrated, which is in agreement with
the small frustration coefficient from the magnetic data. 

\section{M$\text{g}$C$\text{u}$$_{3}$(OH)$_{6}$C$\text{l}$$_{2}$}

\subsection{Synthesis}

\begin{figure}[H]
\noindent \begin{centering}
\includegraphics[width=1\columnwidth]{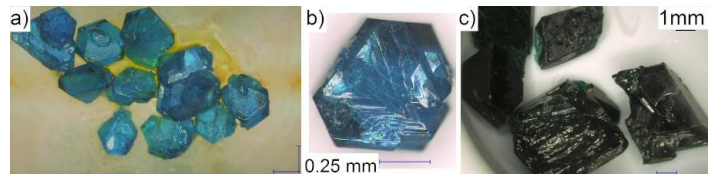}
\par\end{centering}
\caption{\textcolor{black}{\label{Mg}}a), b) Picture of haydeeite crystals
grown in an external gradient. c) Images of the parasitic tondiite
crystals which dominantly grow below 180\textdegree C.}
\end{figure}

\textcolor{black}{So far reports on the synthesis of haydeeite powder
samples were only on a reflux setup with oxygen flushing \citep{haydeeite}.
Here we present a simple way to produce polycrystalline haydeeite
}MgCu$_{3}$(OH)$_{6}$Cl$_{2}$ s\textcolor{black}{amples using a
reaction of 1\,g MgCl$_{2}$$\cdot4$ H$_{2}$O and 1.16\,g CuO
heated to 220\textdegree C for three days in a Parr 4749 acid digestion
vessel. The synthesis is enabled due to the low dissociation temperature
of MgCl$_{2}$$\cdot4$ H$_{2}$O at 110\textdegree C. Single crystals
could then be prepared by the external gradient method using the prereacted
powder as a starting material being dissolved in a solution of 6\,g
MgCl$_{2}$$\cdot4$ H$_{2}$O and 5\,ml distilled water. The solution
was optimized by another pre reaction method of 6\,g MgCl$_{2}$$\cdot4$
H$_{2}$O dissolving 0.48\,g CuO in 5\,ml water at 220\textdegree C
for three days leading up to 100\,$\mu$m haydeeite crystals. It
turned out, that the temperature window for the crystallization of
haydeeite is narrow. We found 180\textdegree C to be the optimal temperature
at the cold end, since the polymorphic tondiite }Mg$_{x}$Cu$_{4-x}$(OH)$_{6}$Cl$_{2}$\textcolor{black}{{}
crystals are formed during the synthesis below 180\textdegree C. These
tondiite crystals grow even up to 100\,mg with 4 x 4 x 3 mm$^{3}$
(see figure \ref{Mg} c). But they have an extremely low substitution
amount of Mg ($0.05<x<0.4$) due to the high MgCl$_{2}$ concentration
being optimized for haydeeite. Similar to herbertsmithite higher MgCl$_{2}$
concentrations lead to a decreased Mg content in the formed tondiite
crystals. With a gradient of 200\textdegree C - 180\textdegree C in
a 15\,cm ampoule we succeeded to grow hexagonal blue single crystals
of up to 0.5 x 0.5 x 0.25\,mm$^{3}$ and 1.6\,mg depicted in figure
\ref{Mg} a, b).}

\subsection{EDS Analysis}

We measured EDS in the SEM on several crystals of different batches
and found a rather high uncertainty for Mg-Cu with this method. To
get a grasp of the Mg content we performed electron probe micro analysis
(EPMA) on up to four crystals of two different batches. We found an
enhanced average substitution content of $x_{Mg}=1.2\pm0.2$, compared
to even stronger enhanced values with $x_{Mg}=1.49\pm0.16$ reported
in Ref. \citep{haydeeite}.

\subsection{Magnetic susceptibility}

\begin{figure}[h]
\noindent \begin{centering}
\includegraphics[width=1\columnwidth]{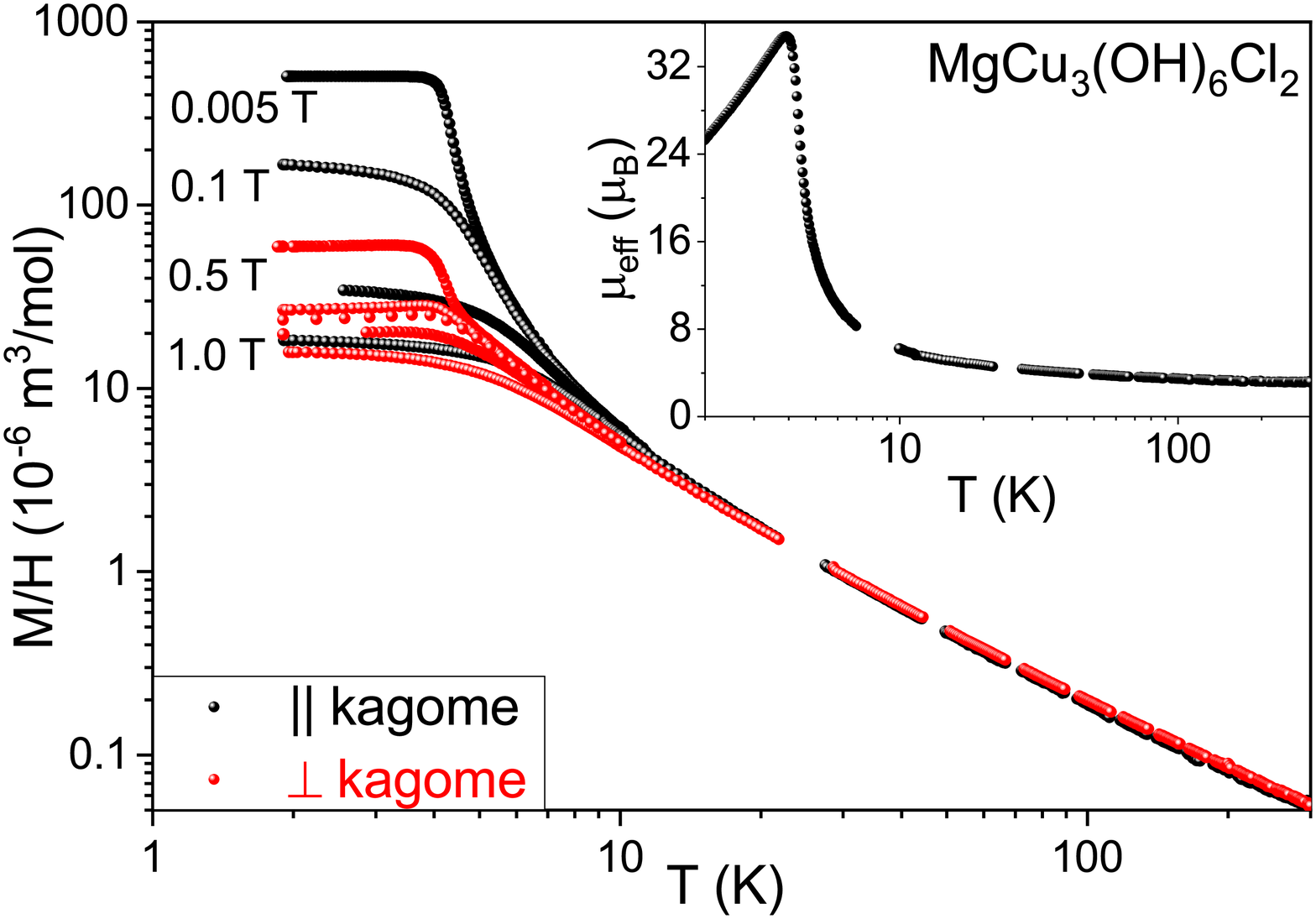}
\par\end{centering}
\caption{\textcolor{black}{\label{susz Mg}The image shows the magnetic data
on a MgCu$_{3}$(OH)$_{6}$Cl$_{2}$ single crystals with a mass of
1.6\,mg of the susceptibility in the range of 1.8 - 300\,K field
cooled at 0.1 - 1\,T with the field aligned parallel (red dots) and
perpendicular to the kagome plane (black dots). The main figure shows
a log-log plot of the temperature dependant susceptibility at various
fields. The inset is a rescaling on the effective moment by a }$\sqrt{\chi T}$
plot of the 50 Oe curve.}
\end{figure}

Magnetic and neutron scattering experiments on polycrystalline haydeeite
revealed a ferromagnetic ground state below T$_{C}$ = 4.2\,K \citep{haydeeite,boldrin(2015)}.
Here, we present the magnetic measurements on a single crystal with
a mass of 0.3\,mg. The susceptibility shown in figure \ref{susz Mg}
was measured both along and perpendicular to the kagome plane and
reveals a small anisotropy below $T_{C}$. The sample shows a ferromagnetic
order around 4.31\,K with a sizeable field dependence stabilizing
the magnetic order as expected for a ferromagnet. From a $\sqrt{\chi T}\sim\mu_{eff}$
plot shown in the inset of figure \ref{susz Mg} the effective moment
value of $\mu_{eff}=$1.9\,$\mu_{B}$ at 290\,K can be derived. 

\begin{figure}[h]
\noindent \begin{centering}
\includegraphics[width=1\columnwidth]{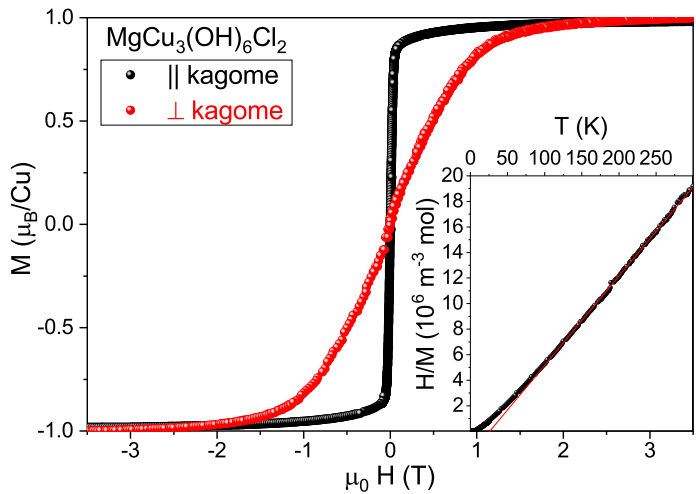}
\par\end{centering}
\caption{\textcolor{black}{\label{M(H)}Field dependant magnetisation measured
at 1.9\,K for both field directions parallel (black) and perpendicular
(red) to the kagome plane. The inset shows the inverse susceptibility
$H/M$ of a stack of single crystals of 1.6\,mg plotted versus temperature
at a field of 0.1\,T applied parallel to the kagome plane.}}
\end{figure}

The magnetization curve in the ordered phase at 1.9 K shown in figure
\ref{M(H)} confirms the ferromagnetic nature of the transition. We
observe an ordered moment of $M_{s}=$1\,\textmu B for both field
directions, with the easy axis within the kagome plane, where the
saturation is reached around 0.1 T. For field perpendicular to the
kagome plane, the critical field is around 2 T. We were not able to
resolve a finite hysteresis curve properly. We note that the ordered
moment is larger than reported in \citep{Colman(2011),boldrin(2015)},
there an ordered moment of 0.28\,\textgreek{m}B (at 500\,Oe) and
$M_{s}=$0.83\,\textgreek{m}B per mole of Cu, was reported. The reason
might be slightly different x values. A Curie-Weiss fit of a stack
of single crystals for a field applied parallel to the kagome plane
is shown on the inset of figure \ref{M(H)} c) and yields Curie-Weiss
temperatures of $\Theta_{W}^{||}=$26\,K compared to $\Theta_{W}=28\pm3$\,K
for powder data \citep{boldrin(2015)}.

\subsection{Specific heat}

\begin{figure}[h]
\noindent \begin{centering}
\includegraphics[width=1\columnwidth]{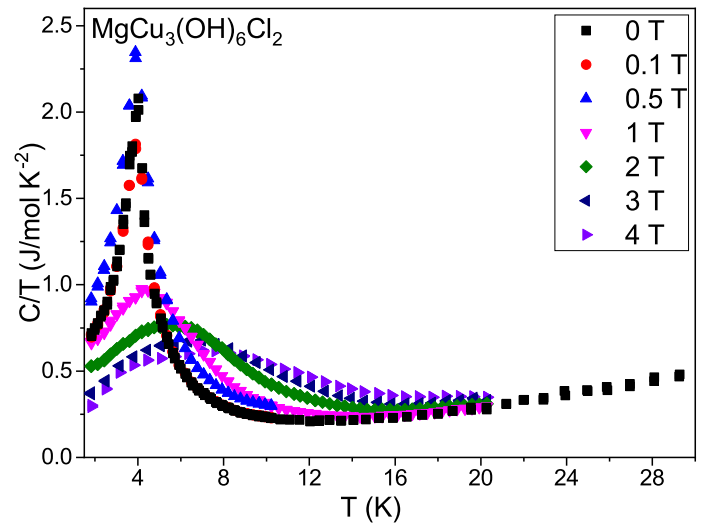}
\par\end{centering}
\caption{\textcolor{black}{\label{HC}Specific heat measurements in an external
field of 0, 0.1, 0.5, 1, 2, 3 and 4\,T applied perpendicular to the
kagome plane, measured from 1.8 to 20\,K on a single crystal of 0.3\,mg.}}
\end{figure}

The observed magnetic transition was also analyzed on several crystals
by specific heat. In figure \ref{HC}, we show a measurement between
1.8\,K and 20\,K on a single crystal of 0.3\,mg. The specific heat
curve divided by temperature is shown in figure \ref{HC} for fields
of 0, 0.1, 0.5, 1, 2, 3 and 4\,T along the kagome plane. We roughly
estimated the contributions in a temperature range of 15 - 32\,K
from a linear fit of a C/T vs T$^{2}$ plot. With $\beta\approx0.402(7)$\,mJ/molK$^{-2}$
we get an estimated debye temperature of $\Theta_{D}\approx443$\,K.
We observe a sharp peak at the magnetic transition, which can best
be seen in the $C/T$ plot in figure \ref{HC} which shows a strong
field dependance shifting it up to 10\,K at 4\,T. The increased
area under the peak compared to the two previously shown compounds
is in agreement with the full captured moment in the $M(H)$ curve.

\section{Conclusion}

In conclusion we have presented results on three members of two-dimensional
kagome quantum spin systems in the atacamite family. The successful
hydrothermal synthesis of single crystals for these three candidates
has been shown. They reveal very different magnetic ground states.
EuCu$_{3}$(OH)$_{6}$Cl$_{3}$ is a kapellasite-like kagome material
with the highest reported Curie-Weiss temperature for the atacamite
family of -400\,K and presents the strongest anisotropy in this family
of up to 2.7 due to Van Vleck paramagnetic Eu$^{3+}$ contributions.
The system is strongly frustrated with partial order at 14.7\,K and
a spin reorientation at 8.7\,K. The frustration coefficient is around
30. ZnCu$_{3}$(OH)$_{6}$(NO$_{3}$)$_{2}$ turned out to be an antiferromagnetic
system without strong frustration effects. Magnetic measurements on
single crystals revealed only tiny anisotropy in this compound. Furthermore,
we present measurements on single crystals of haydeeite and confirmed
the ferromagnetic order at 4.3\,K measured previously on polycrystals.
We detect a small but well-defined anisotropy in the ordered phase
with the easy axis within the kagome plane. In addition the specific-heat
measurements reveal a clear anomaly, confirming the absence of sizeable
frustration effects. 

These results together with a compilation of the physical and structural
properties of kagome materials around the atacamite family shows us
a rich playground where the perfect spin liquid candidate is in close
reach. The clue to the latter might be in a more in-depth study of
the delicate interplay of the different magnetic exchange couplings.
The materials presented in this study give a firm basis to study this
influence on the overall degree of frustration in kagome quantum spin
systems. Finally, from the presented kagome systems and the listed
ones one can assume that an increasing seperation of the kagome plane
does not stabilize the spin liquid properties.
\begin{acknowledgments}
The authors gratefully acknowledge support by the Deutsche Forschungsgemeinschaft
through grant SFB/TR 49.
\end{acknowledgments}

\end{document}